\documentclass[12pt]{article}
\usepackage[top=1.18in, left=1.1in, right=1.1in, bottom=1in, includefoot]{geometry}
\usepackage{array}
\usepackage{booktabs}
\usepackage{comment}

\usepackage[svgnames]{xcolor}
\usepackage{rotating}
\usepackage{amsmath}
\usepackage{graphicx}

\usepackage{undertilde}
\usepackage[T1]{fontenc}
\usepackage{titlesec,blindtext}
\usepackage{enumitem}
\usepackage{graphicx}
\usepackage{placeins}
\usepackage{setspace}
\usepackage{natbib}
\usepackage{listings}
\usepackage{datetime}

\usepackage{algorithm}

\usepackage[noend]{algpseudocode}

\newcommand{\bmath}{\boldsymbol}

\titleformat{\section}{\large\bfseries}{\thesection}{1em}{}
\titleformat{\subsection}[block]{\hspace{1.5em}\normalsize\bfseries}{\thesubsection}{1em}{}
\titleformat{\subsubsection}[block]{\hspace{2.5em}\normalsize\bfseries}{\thesubsubsection}{1em}{}
\titleformat{\paragraph}[block]{\hspace{3.5em}\normalfont\normalsize\bfseries}{\theparagraph}{1em}{}

\def\modif#1{#1}
\begin{document}

\begingroup  
  \centering
  \large \textbf{Simple fixed-effects inference for complex functional models}\\[0.5em]
  \large SY Park$\null^{1}$, AM Staicu$\null^{1}$, L Xiao$\null^{1}$, and CM Crainiceanu$\null^{2}$ \\ [0.2em]
  \normalsize $\null^{1}$Department of Statistics, North Carolina State University\\[0.2em] 
    \normalsize $\null^{2}$Department of Biostatistics, School of Public Health, Johns Hopkins University\\[0.3em]  
  \today \\[1em]
\endgroup
%\doublespacing

\begin{abstract}
We propose simple inferential approaches for the fixed effects in complex functional mixed effects models. We estimate the fixed effects under the independence of functional residuals assumption and then bootstrap independent units (e.g. subjects) to estimate the variability of and conduct inference in the form of hypothesis testing on the fixed effects parameters. Simulations show excellent coverage probability of the confidence intervals and size of tests. Methods are motivated by and applied to the Baltimore Longitudinal Study of Aging (BLSA), though they are applicable to other studies that collect correlated functional data.

\end{abstract}

{\sc Keywords:} Bootstrap/resampling, Functional data, Measurement error, Smoothing and nonparametric regression. 

\section{Introduction} 
Rapid advancement in technology and computation has led to an increasing number of studies that collect complex-correlated functional data. In response to these studies research in structured functional data analysis (FDA) has witnessed rapid development. A major characteristic of these data is that they are strongly correlated, as multiple functions are observed on the same observational unit. Many new studies have functional structures including multilevel \citep{di2009multilevel, crainiceanu2009generalized}, longitudinal \citep{greven2010,chen2013repeated, scheipl2014functional}, spatially aligned \citep{baladandayuthapani2008bayesian, staicu2010fast, serban2013biometrics}, or crossed \citep{morris2006wavelet, aston2010linguistic, zhu2011robust, shou2013structured}.

While these types of data can have highly complex dependence structures, one is often interested in simple, population-level, questions for which the multi-layered structure of the correlation is just an infinite-dimensional nuisance parameter. For example, in the Baltimore Study of Aging (BLSA) activity data are collected for each participant at the minute level for multiple days. Thus, data exhibit complex within-day (the circadian rhythm of daily activity) and between-day (the circadian rhythm of activity across days for the same subject) correlations. However, the most important questions in the BLSA tend to be simple; in particular, one may be interested in how age affects the circadian rhythm of activity or whether the effect is different by gender. In this context the high complexity and size of the data are just technical inconveniences.

Such simple questions are typically answered by estimating fixed effects in complex functional mixed effects models. Our alternative proposal avoids the heavy associated computations by: 1) estimating the fixed (population-level) effects under the assumption of independence of functional residuals; and 2) using a nonparametric bootstrap of independent units (e.g. subjects) to construct confidence intervals and conduct tests. A natural question is whether efficiency is lost by ignoring the correlation. While the loss of efficiency is well documented in longitudinal studies with few observations per subject and small dimensional within-subject correlation, little is known about inference when there are many observations per subject with an unknown large dimensional within-subject correlation matrix. Our own view is that estimating large dimensional covariance matrices of functional data to estimate fixed effects may actually waste degrees of freedom. Indeed, a covariance matrix for an $n $ by $p$ matrix of functional data ($n$ = number of subjects and $p$ = number of subject-specific observations) would require estimation of $p(p+1)/2$ matrix covariance entries. When $p$ is moderate or large and the covariance matrix is unstructured this is a difficult problem. Moreover, the resulting matrix has an unknown low rank and is not invertible. 

%Correlated functional data has received great attention lately: \citet{morris2006wavelet}; \citet{}; \citet{di2009multilevel}; \citet{staicu2010fast}; \citet{greven2010}; \citet{}; \citet{}; \citet{} \citet{}; \citet{li2014fpca} to name a few. These methods divide into two main categories, based on whether they account for additional covariates or not. While they mainly focus on modeling of the complex data, the development of inferential procedures for the covariate effect has received less attention. This paper is concerned with developing inferential procedures in the form of confidence intervals and testing procedures, for the covariate effects in functional regression models when the response is correlated functional data.

We will consider cases when multiple functional observations are observed for the same subject. This structure is inspired by many current observational studies, but we will focus on the BLSA, where activity data are recorded at the minute level over multiple days, resulting in daily activity profiles observed over multiple days. Consider that the observed data is of the form $\{ Y_{ij}(\cdot), \bmath{X}_{ij} \}$, where $Y_{ij}(\cdot)$ is the $j$th unit functional response (e.g. $j$th visit) for the $i$th subject, and $\bmath{X}_{ij}$ is the corresponding vector of covariates. This general form applies to all types of functional data discussed above: multilevel, longitudinal, spatially-correlated, crossed, etc. The main objective is to make statistical inference for the population-level effects of interest. 

% Let $\mu(t,\bmath{X}_{ij})$ be the mean function of interest, where $t$ is the functional argument. The model we are interested in is
%  $$Y_{ij}(t)=\mu(t,\bmath{X}_{ij})+  \epsilon_{ij}(t)$$
%where $\epsilon_{ij}(t)$ can have complex correlation structures induced, for example, by exchangeable or spatial clustering of functional data.
%We are interested in producing an estimator of $\mu(t,\bmath{X}_{ij})$, $\widehat{\mu}(t,\bmath{X}_{ij})$, together with an estimator of its variability. The goal is to conduct inference on $\mu(t,\bmath{X}_{ij})$.

One na\"ive approach to analyze data with such complex structure is to ignore the dependence over the functional argument $t$, but to account for the dependence across the repeated visits; specifically by assuming that responses $Y_{ij}(t)$ are correlated over $j$ and independent over $t$. Longitudinal data analysis literature offers a wide variety of models and methods for estimating the fixed effects and their uncertainty, and for conducting tests (see for example \cite{laird1982random, liang1986longitudinal, fitzmaurice2012applied}). These methods allow to account for within-subject correlation, incorporate additional covariates, and make inference about the fixed effects. Nevertheless, extending these estimation procedures to functional data is difficult because specifying the dependence for functional data is not obvious while implementation may be very computationally expensive.

Another possible approach is to completely ignore the dependence across the repeated visits $j$, but account for the functional dependence; specifically assume $Y_{ij}(t)$ are dependent over $t$, but independent over $j$. Function on scalar/vector regression models can be used to estimate the fixed effects of interest; see for example  \citet{faraway1997regression,jiang2011functional, ivanescu2013penalized}. In this context, testing procedures for hypotheses on fixed effects are available. For example, \citet{shen2004f} proposed the functional F statistic for testing hypotheses related to nested functional linear models. \citet{zhang2007statistical} proposed $L_2$ norm based test for testing the effect of a linear combination of time-varying coefficients, and approximate the null sampling distribution using resampling methods. However failing to account for all sources of dependence results in tests with inflated type I error.

In contrast, development of statistical inference methods for correlated functional data has received less attention. For example,  \citet{morris2006wavelet} discussed Bayesian inference in the functional mixed model framework; however, their main focus was on modeling, and hypothesis testing was not studied. \citet{CMC2011} discussed bootstrap-based inferential methods for the difference in the mean profiles. \citet{staicu2014likelihood} proposed likelihood-ratio type testing procedure, while \citet{staicu2014significance} considered $L_2$ norm-based testing procedures for testing the null hypothesis that multiple group mean functions are equal. \citet{horvath2013estimation} developed inference for the mean function of a functional time series. Nevertheless, none of these papers handle inference on fixed effects in full generality. Here we consider a modeling framework that is a direct generalization of the linear mixed model framework from longitudinal data analysis, where scalar responses are replaced with functional ones. We study confidence intervals and testing procedures for the fixed effects using bootstrap methods over subjects to account for all known sources of data dependence.  
  
The rest of the paper is organized as follows. Section \ref{sec:model} introduces the modeling and estimation framework and discusses several important examples. Section \ref{sec:variability} describes an approach to quantifying the variability of the estimators using bootstrap. Section \ref{sec:testStat} proposes formal test procedure for the null hypothesis that the mean function does not depend on a covariate of interest. Applications and simulation results are presented in Sections \ref{sec:blsa} and \ref{sec:simulation}, respectively. We conclude with a brief discussion in Section \ref{sec:discussion}.

\section{Modeling framework} \label{sec:model}

Consider the case when each subject is observed at $m_i$ visit times, and data at each visit consist of a functional outcome $\{Y_{ij\ell}=Y_{ij}(t_{ij\ell}): \ell=1, \ldots, L_{ij} \} $ and a vector of covariates including a scalar covariate of interest, $X_{ij}$, and additional $p$-dimensional vector of covariates, $\mathbf{Z}_{ij}$. We assume that $t_{ij\ell} \in \mathcal{T}$ for compact and closed domain $\mathcal{T}$; take $\mathcal{T}=[0,1]$ for simplicity. For convenience, we assume a balanced regular sampling design, i.e. $t_{ij\ell} = t_\ell$ and $L_{ij}=L$, though all methods apply to general sampling designs. Furthermore, we assume that $\{X_{ij}: \forall \; i,j\}$ is  a dense set in the closed domain $\mathcal{X}$\modif{; this assumption is needed for the case when the fixed effect of $X_{ij}$ is modeled nonparametrically \citep{ruppert2003semiparametric, fitzmaurice2012applied}. }A common approach for the study of the effect of the covariates on the functional outcome $Y_{ij}(\cdot)$ is to posit a model of the type
\begin{equation}
Y_{ij}(t) = \mu(t, X_{ij}) + \mathbf{Z}^T_{ij}\boldsymbol{\tau} +  \epsilon_{ij}(t), \label{model:initial}
\end{equation}
where $\mu(t, X_{ij})$ is a time-varying smooth effect of the covariate of interest, $X_{ij}$, and $\boldsymbol{\tau}$ is a $p$-dimensional parameter quantifying the linear additive effect of the covariate vector, $\mathbf{Z}_{ij}$. Here $ \epsilon_{ij}(t)$ is a zero-mean random deviation that incorporates both the within- and between-subject variability. Below we present several examples of models for $\mu(t,X_{ij})$ that are relevant to our problem: 
\begin{enumerate} [label= \ref{sec:model}(\alph*)]
\item  $\mu(t, X_{ij}) = \beta_0 + \beta_tt + \beta_xX_{ij}$ \label{ex:linear}
\item  $\mu(t, X_{ij}) = \beta_0 + \beta_tt + \beta_xX_{ij} + \beta_{tx}tX_{ij}$ \label{ex:interaction}
\item  $\mu(t, X_{ij}) = f(t) + \beta_xX_{ij}$, where $f(\cdot)$ is an unknown smooth function \label{ex:univariate} 
%\item  $\mu(t, X_{ij}) = f(t) + g(X_{ij})$, where $f(\cdot)$ and $g(\cdot)$ are unknown univariate smooths \label{ex:additive}
\item  $\mu(t, X_{ij}) = h(t,X_{ij})$, where $h(\cdot, \cdot)$ is an unknown bivariate smooth function \label{ex:nonparametric}
\end{enumerate}
\noindent Models \ref{ex:linear} and \ref{ex:interaction} assume a linear effect of both the functional argument, $t$, and the covariate of interest, $X_{ij}$, with or without interaction effects. In particular, model \ref{ex:linear} assumes that the rate of change of the mean response with respect to $t$ is constant and does not depend on $X_{ij}$, while model \ref{ex:interaction} assumes that the rate of change depends on the covariate of interest.
Model \ref{ex:univariate} describes an additive effect of the functional argument and covariate of interest, with the additional assumption that the effect of $X_{ij}$ is linear. The mean model \ref{ex:nonparametric} describes a completely nonparametric structure. While this model is useful when there is no a priori information on the mean structure, fitting a nonparametric bivariate function is computationally expensive. We considered the case when $X_{ij}$ is univariate mainly to keep the number of indices under control. All methods can be applied in more generality.

Fitting model (\ref{model:initial}) with either of the mean structures \ref{ex:linear}-\ref{ex:nonparametric} is not new. \cite{morris2006wavelet}, and \cite{scheipl2014functional} discuss estimation of the mean parameters in a variety of cases using an independence working assumption across observations. Also, when $X_{ij}$ is the actual visit time, and there are no other covariates is the study, then the approach in \cite{chen2013repeated} can be used to estimate a bivariate smooth mean under the working independence assumption. However, none of these papers discusses inference on the population level effects that accounts for the complex correlation structure of the data. The novelty of this paper consists precisely in filling this gap in the literature. To be specific, we consider an estimation approach based on the independence working assumption, introduce pointwise and joint confidence bands for the fixed effects, and propose a hypothesis testing procedure for the null hypothesis that the covariate of interest, $X_{ij}$, has no effect on the outcome.   

When the data are modeled as in model (\ref{model:initial}) and $\mu(t,X)$ has the structure \ref{ex:linear}, then the mean parameter estimates are $\beta_0, \beta_t$, $\beta_x$, and $\boldsymbol{\tau}$. They are estimated by minimizing $SSE = \sum_{i,j,\ell}[Y_{ij\ell} - \{\beta_0 +\beta_t t_{ij\ell}+ \beta_x X_{ij}+ \mathbf{Z}^T_{ij}\boldsymbol{\tau} \} ]^2$. Estimators can be represented in matrix form as  $[\boldsymbol{\widehat \beta}^T, \boldsymbol{\widehat \tau}^T]^T = (\mathbf{M}^T\mathbf{M})^{-1}\mathbf{M}^T\mathbf{Y}$, where $\boldsymbol\beta=(\beta_0, \beta_t, \beta_x)^T$, $\mathbf{M} = [\mathbf{M_1}\text{ } \vdots\text{ } \mathbf{M_2}]$, with $\mathbf{M_1}$ the matrix with rows $(1,t_{ij\ell}, X_{ij})$ and $\mathbf{M_2}$ the matrix obtained by row-stacking of $ \mathbf{Z}^T_{ij}$. Here $\mathbf{Y}$ is the $L \ \sum_{i=1}^n m_i$- dimensional vector of all $Y_{ij\ell}$'s.

The estimation criterion becomes progressively more complicated as the mean structure $\mu(t,X)$ becomes more involved. For example, for nonparametric modeling (\ref{ex:univariate} and \ref{ex:nonparametric}) we follow standard smoothing practices using penalized splines. Of course, other types of smoothers are also acceptable and may be equally or more appropriate for different types of data structures. The methods we discuss apply to all types of smoothers. To be specific, consider the most complex example, \ref{ex:nonparametric}, where $\mu(t,X)$ is an unspecified bivariate smooth function. Construct a bivariate basis by the tensor product of two univariate B-spline bases, $\{B^{t}_{1}(t),\cdots,B^{t}_{d_t}(t)\}$, and $\{B^{x}_{1}(x),\cdots,B^{x}_{d_x}(x)\}$, defined on $\mathcal{T}$ and $\mathcal{X}$ respectively. Then $\mu(t,x)$ $= \sum_{l=1}^{d_t}\sum_{r=1}^{d_x} B^{t}_l(t)B^{x}_r(x) \beta_{lr}={\mathbf{B}(t,x)^T} {\boldsymbol{\beta}}$; where $\mathbf{B}(t,x)$ is the $d_t d_x$-dimensional vector of $B^{t}_l(t)B^{x}_r(x)$'s and ${\boldsymbol{\beta}}$ is the vector of parameters $\beta_{lr}$. Typically, the number of basis functions is chosen sufficiently large to capture the maximum complexity of the mean function and smoothness is induced by a quadratic penalty on the coefficients. There are several penalties for bivariate smoothing with the most popular being the ones proposed by \cite{marx2005multidimensional} and  \cite{wood2006, wood2006a}. More recently,  \cite{xiao2013fast, xiao2013FACE} proposed a scalable sandwich penalty estimator that leads to a computationally efficient algorithm for high dimensional data. In this paper we used the following estimation criterion 

%One problem with these penalties is that the associated algorithms are slow and do not generalize well to high dimensions. \cite{xiao2013fast, xiao2013FACE} proposed a scalable sandwich penalty estimator that can be used in our activity study. 
% introduced roughness penalties to measure the roughness of a function; $PEN = \lambda_t\int \int \{\partial^2 \mu(t,x)/\partial t^2 \}^2 dt dx + \lambda_x\int \int \{\partial^2\mu(t,x)/\partial x^2 \}^2 dt dx  $, where $\lambda_t$ and $\lambda_x$ control the amount of smoothness in directions $t$ and $x$ respectively. This penalty can be further represented as $\boldsymbol{\beta}^T(\lambda_t \mathbf{P_t} \otimes \mathbf{I_{d_x}} + \lambda_x \mathbf{I_{d_t}} \otimes \mathbf{P_x})\boldsymbol{\beta}$, where $ \mathbf{P_t}$ and $ \mathbf{P_x}$ are $d_t\times d_t$ and $d_x\times d_x$ square matrices with elements given by $\int B^{\prime\prime}_{t,l}(t) B^{\prime\prime}_{t,l'}(t) dt $ and $\int B^{\prime\prime}_{x,r}(x) B^{\prime\prime}_{x,r'}(x) dx % $ respectively (see \cite{wood2006}). 
%Xiao's penalty has the familiar quadratic form
%
\begin{equation} \label{eq:objectiveFn}
\underset{\boldsymbol{\beta}, \text{  } \boldsymbol{\tau}, \text{ } \lambda}{\text{argmin}} \text{  } \sum_{i, j, \ell} [Y_{ij\ell}-\{\mathbf{B}(t_\ell,X_{ij})^T\boldsymbol{\beta}+\mathbf{Z}^T_{ij}\boldsymbol{\tau}\} ]^2 + \boldsymbol{\beta}^T
%(\lambda_t \mathbf{P_t} \otimes \mathbf{I_{d_x}} + \lambda_x \mathbf{I_{d_t}} \otimes \mathbf{P_x})
P_\lambda\boldsymbol{\beta},
\end{equation} 
with a penalty matrix $P_\lambda$ described in \cite{wood2006} and a vector of smoothing parameters, $\lambda$. \modif{Specifically, we used $P_{\lambda} = \lambda_t P_{t} \otimes I_{d_x} + \lambda_x I_{d_t} \otimes P_{x}$ and $\lambda = (\lambda_{t}, \lambda_x)^{T}$, where $\otimes$ denotes the tensor product, and $P_t$ and $\lambda_t$ are the marginal second order difference matrix and the smoothing parameter for the $t$ direction, respectively; $P_x$ and $\lambda_x$ are defined similarly for the $x$ direction. Here $I_{d_t}$ and $I_{d_x}$ are the identity matrices of dimensions $d_t$ and $d_x$.} 
%The intuition for the scalability of Xiao's penalty is that it is chosen so that it ``completes the sum of squares'', whereas all other penalties do not. More precisely, in the familiar formula $(a+b)^2=a^2+2ab+b^2$ one would need to add the $2ab$ term to the $a^2+b^2$ term to simplify the formula. That is exactly what Xiao's penalty does, which makes it scalable not only in terms of sample and dimension size, but also in terms of the dimension of the function, $\mu(t,x)$. 
For a fixed smoothing parameter, $\lambda$, the minimizer of (\ref{eq:objectiveFn}) has the form $[\boldsymbol{\widehat \beta}_\lambda^T, \boldsymbol{\widehat \tau}_\lambda^T]^T  = (\mathbf{M}^T\mathbf{M}+P_\lambda)^{-1}\mathbf{M}^T\mathbf{Y}
$, while the estimated mean is 
$\widehat{\mu}(t,x) + \mathbf{Z}^T_{ij}\widehat{\boldsymbol{\tau}} = \mathbf{B}(t,x)^T \boldsymbol{\widehat{\beta}}_\lambda + \mathbf{Z}^T_{ij}\widehat{\boldsymbol{\tau}}_\lambda$.

%KEEP this: $\underset{d_td_x \times 1}{\mathbf{B}(t,x)^T}\underset{d_td_x \times 1}{\boldsymbol{\beta}}$. To     
Selecting the optimal value of the smoothing parameter has been discussed extensively in the literature. Two widely used criteria  are the generalized cross validation (GCV) and the restricted maximum likelihood (REML). GCV is based on prediction error, whereas REML is based on likelihood estimation where the a smoothing parameter is a variance parameter. Empirical evidence \citep{ruppert2003semiparametric} suggests that REML and GCV tend to have different behaviors because of the different way they trade bias for variance. REML tends to be more biased with lower variance (\cite{ruppert2003semiparametric, reiss2007, reiss2009, wood2006}), while GCV tends to be less biased with higher variance (\cite{ruppert2003semiparametric, wahba1990spline}). New evidence (\cite{xiao2014quantifying}) suggests that covariance smoothing can be improved by using leave-one-subject-out generalized cross validation for functional data. However, here we investigate only estimation under independence both for the mean function and its smoothing parameters\modif{; in our numerical investigation we select the optimal smoothing parameters by GCV.}

%We consider REML in our numerical investigation. This choice is motivated primarily by the advantage of REML in terms of numerical stability that has been reported frequently in literature (see \citet{reiss2007}, \citet{reiss2009}, and \cite{wood2011faststable}). Also our choice is  motivated by the theoretical properties of the REML based selection with regards to robustness to moderate violations of the independent error assumption, see \cite{krivobokova2007note}.

The \verb|gam| function in the \verb|R| \citep{R} package \verb|mgcv| \citep{wood2006} is used to implement model (\ref{model:initial}) with various mean structures as discussed above and using row and column penalties. The \verb|fbps| function \citep{xiao2013fast} in \verb|R| \citep{R} package \verb|refund| \citep{Rrefund} can be used to fit the smooth bivariate effects by employing fast bivariate penalized spline smoothing with a modified penalty \citep{xiao2013FACE}. This function requires only simple modifications to account for linear effects of additional covariates: the linear effects are fit first and the bivariate smooth effects are fitted second conditional on the linear effects estimates. The reason for considering a range of models from very simple parametric to complex nonparametric models is to show the generality of the approaches. While we will keep a close eye on the activity application, the basic principle remains simple: estimate parameters under independence and bootstrap independent units.

\begin{comment}
We use cubic penalized spline bases for both $t$ and $x$ directions with a total dimension of $25$ (the marginal dimensions $d_t$ and $d_x$ being equal to $5$ each). If a simpler model is fitted, then adjustments need to be made on both the basis function and the roughness penalty. Specifically, for Examples \ref{ex:linear} and \ref{ex:interaction}, $\mathbf{B}(t,x)=[1,t,x]^T$ and $\mathbf{B}(t,x)=[1,t,x,tx]^T$, respectively, with no penalty in Equation (\ref{eq:objectiveFn}). For Example \ref{ex:univariate}, $\mathbf{B}(t,x)=[B^{t}_1(t),...,B^{t}_{d_t}(t),x]^T$, with the penalty $ \mathbf{P_1}=\lambda_t[\{\mathbf{P_t}^T \text{ }\vdots\text{ } \mathbf{0}_{d_t \times 1} \}^T \text{ }\vdots\text{ } \mathbf{0}_{ (d_t+1) \times 1} ]$. Lastly, for Example \ref{ex:additive}, $\mathbf{B}(t,x)=[B^{t}_1(t),...,B^{t}_{d_t}(t),B^{x}_1(x),...,B^{x}_{d_x}(x)]^T$, with the penalty term $\mathbf{P_1}=(\lambda_t[\{\mathbf{P_t}^T \text{ }\vdots\text{ } \mathbf{0}^T_{(d_x \times d_t)}\}^T \text{ }\vdots \text{ } \mathbf{0}_{( (d_t+d_x) \times d_x)}] + \lambda_x[\mathbf{0}_{( (d_t+d_x) \times d_t)}\text{ }\vdots\text{ }\{\mathbf{0}^T_{(d_t \times d_x)} \text{ }\vdots\text{ } \mathbf{P_x}^T\}^T])$.

\end{comment}

In the following section we discuss inference for $\mu(t,x)$ in the form of confidence bands and hypothesis testing. 

\section{Confidence bands for $\mu(t,x)$}\label{sec:variability}

Without loss of generality, assume that the mean structure is $\mu(t,x) =  \mathbf{B}(t,x)^T \boldsymbol{{\beta}}$, where $ \mathbf{B}(t,x)^T$ can be as simple as $(1,t,x)$ or as complex as a vector of pre-specified basis functions. The mean estimator of interest is $\widehat{\mu}(t,x) = \mathbf{B}(t,x)^T \boldsymbol{\widehat{\beta}}$. One could study pointwise variability for every pair $(t,x)$, that is $\textrm{var}\{\widehat \mu(t,x)\}$, or the joint variability for the entire domain $\mathcal{T}\times \mathcal{X}$, that is $\textrm{cov}\{\widehat \mu(t,x): t\in \mathcal{T}, x\in \mathcal{X}\}$. Irrespective of the choice, the variability is fully described by the variability of the parameter estimator $\boldsymbol{\widehat{\beta}}$.

In this paper we consider a flexible dependence structure for $\epsilon_{ij}(t)$ that describes both within- and between-subject variability. \modif{We make minimal assumptions on the errors that $\epsilon_{ij}(t)$ is independent over $i$ but is correlated over $j$ and $t$.} Deriving the analytical expression for the sampling variability of the estimator $\boldsymbol{\widehat{\beta}}$ in such contexts is challenging. We use bootstrap to study the sampling properties of the parameter estimator. Two bootstrap algorithms are discussed: bootstrap of subject-level data and bootstrap of subject-level residuals. \modif{These approaches have already been studied and popularly used under the nonparametric regression setting for independent measurements; see, for example, \citet{hardle1988bootstrapping}, \citet{efron1994introduction}, and \citet{hall2013simple} among many others. Bootstrap of functional data for fixed effects has also been considered in several literatures, including \citet{politis1994stationary} for weakly dependent processes in Hilbert space, \citet{cuevas2006use} for independent functional data, and \citet{crainiceanu2011statistical} for two paired samples of functional data. Nonetheless, performance of the proposed bootstrap algorithms for dependent functional data with such complex error structures that we consider in this paper is unknown and needs to be assessed.}  

% \modif{These approaches have been studied under the regression setting for various data structures, including independent measurements, weakly dependent time-series, independent functional data, and etc.; see, for example, \citet{hardle1988bootstrapping, efron1994introduction, hall2013simple, politis1994stationary, cuevas2006use} among many others. Nonetheless, two approaches have not yet been considered for dependent functional data and their performances need to be investigated. }

The first method is more generally applicable, while the second relies on two important assumptions: i) the covariates do not depend on visit, that is $X_{ij}=X_i$ and $\mathbf{Z}_{ij}=\mathbf{Z}_i$; and ii) both the correlation and the variance of errors are independent of the covariates. 
\modif{These assumptions ensure that sets of subject-level errors, i.e. $\{\epsilon_{ij}(t): j = 1,\ldots, m_i\}$ for $i = 1, \ldots, n$, can be re-sampled over subjects without affecting the sampling distribution.}
Both bootstrap methods rely on specification of $\mathbf{B}(t,x)$. In models that require smoothing parameters, their selection is considered to be part of the estimation procedure and is repeated at each bootstrap step.

% THIS uses the following packages: \usepackage{algorithm}; \usepackage[noend]{algpseudocode}

\begin{algorithm}[H]
\caption{Bootstrap of the subject-level data [uncertainty estimation]} 
\label{sec:bootstrapSub}
\begin{algorithmic}[1]
\For{$b \in \{1, \ldots, B\}$}
    \State Re-sample the subject indexes from the index set $\{1,\ldots,n\}$ with replacement. Let $I^{(b)}$ be the resulting sample of $n$ subjects.
    \State Define the $b$th bootstrap data by: \\
    [] $\text{{data}}^{(b)} =[\{Y_{i^*j}(t_\ell), X_{i^*j}, \mathbf{Z}_{i^*j}, t_\ell\}:i^* \in {I}^{(b)},\text{ } j = 1,\ldots, m_{i^*}, \text{ and } \ell=1,\ldots,L ]$.
    \State  Using $\text{{data}}^{(b)}$ fit the model (\ref{model:initial}) with the mean structure of interest modeled by $\mu(t,x)= \mathbf{B}(t,x)^T \boldsymbol{{\beta}}$, by employing criterion (\ref{eq:objectiveFn}). Let $ \boldsymbol{{\widehat \beta}}^{(b)}_{\lambda}$ be the corresponding estimate of the parameter of interest; similarly define $\widehat \mu^{(b)}(t,x) = \mathbf{B}(t,x)^T\boldsymbol{{\widehat \beta}}^{(b)}_{\lambda^{(b)}}$.
\EndFor
\State Calculate the sample covariance of $\{\boldsymbol{{\widehat\beta}}^{(1)}_{\lambda^{(1)}}, \ldots, \boldsymbol{{\widehat \beta}}^{(B)}_{\lambda^{(B)}}\}$; denote it by $V_{\boldsymbol{{\widehat \beta}}}$.
\end{algorithmic}
\end{algorithm}

In many applications covariates do not depend on the visit (e.g. gender, age), that is $X_{ij}=X_i$ and $\mathbf{Z}_{ij}=\mathbf{Z}_i$; in particular, this is the case in the BLSA data. To account for this information we propose another version of the bootstrap of the data, which relies on the assumption that the error covariance is independent of the covariates. The bootstrap of subject-level residuals shows excellent numerical results, as illustrated in the simulation section. 

Fit the model (\ref{model:initial}) with the mean structure of interest modeled by $\mu(t,x)= \mathbf{B}(t,x)^T \boldsymbol{{\beta}}$, by employing the estimation criterion described in (\ref{eq:objectiveFn}). Calculate residuals by $e_{ij}(t_\ell) = Y_{ij}(t_\ell) -\mathbf{B}(t_\ell,X_{i})^T \boldsymbol{{\widehat \beta}}_{\lambda}- \mathbf{Z}^T_{i}\widehat{\boldsymbol{\tau}}_{\lambda}$.

\begin{algorithm}[H]
\caption{Bootstrap of the subject-level residuals [uncertainty estimation]} 
\label{sec:bootstrapResid}
\begin{algorithmic}[1]
\For{$b \in \{1, \ldots, B\}$}
    \State Re-sample the subject indexes from the index set $\{1,\ldots,n\}$ with replacement. Let $I^{(b)}$ be the resulting sample of subjects. For each $i=1, \ldots, n$ denote by $m^*_i$ the number of repeated time-visits for the $i$th subject selected in $I^{(b)}$.
     
    \State Define the $b$th bootstrap sample of residuals \\
    [] $\{e^*_{ij}(t_\ell):i=1, \ldots, n, \text{ } j = 1,\ldots, m^*_{i}, \text{ and } \ell=1,\ldots,L \}$.
    \State Define the $b$th bootstrap data by: \\
    [] $\text{{data}}^{(b)} =[\{Y^*_{ij}(t_\ell), X_i, Z_i, t_\ell\}:i=1, \ldots, n, j = 1,\ldots, m^*_{i}, \ell=1,\ldots,L ]$, where $Y^*_{ij}(t_\ell)   
    =\mathbf{B}(t_\ell,X_{i})^T \boldsymbol{{\widehat \beta}}_{\lambda}+ \mathbf{Z}^T_{i}\widehat{\boldsymbol{\tau}}_{\lambda} + e^*_{ij}(t_\ell)  :i=1, \ldots, n, j = 1,\ldots, m^*_{i}, \ell=1,\ldots,L \}$.
    \State  Using $\text{{data}}^{(b)}$ fit the model (\ref{model:initial}) with the mean structure of interest modeled by $\mu(t,x)= \mathbf{B}(t,x)^T \boldsymbol{{\beta}}$, by employing criterion (\ref{eq:objectiveFn}). Let $ \boldsymbol{{\widehat \beta}}^{(b)}$ be the corresponding estimate of the parameter of interest; similarly define $\widehat \mu^{(b)}(t,x) = \mathbf{B}(t,x)^T\boldsymbol{{\widehat \beta}}^{(b)}_{\lambda^{(b)}}$.
\EndFor
\State Calculate the sample covariance of $\{\boldsymbol{{\widehat\beta}}^{(1)}_{\lambda^{(1)}}, \ldots, \boldsymbol{{\widehat \beta}}^{(B)}_{\lambda^{(B)}}\}$; denote it by $V_{\boldsymbol{{\widehat \beta}}}$.
\end{algorithmic}
\end{algorithm}

For fixed $(t,x)$, the variance of the estimator $\widehat \mu(t,x) =  \mathbf{B}(t,x)^T \boldsymbol{{\widehat \beta}}$ can be estimated as $\textrm{var}\{\widehat \mu(t,x)\} = \mathbf{B}(t,x)^T \  V_{\boldsymbol{{\widehat \beta}}} \ \mathbf{B}(t,x)$, by using the bootstrap-based estimate of the covariance of $\boldsymbol{{\widehat \beta}}$. A $100(1-\alpha)\%$ pointwise confidence interval for $ \mu(t,x)$ can be calculated as $\widehat \mu(t,x) \pm z^*_{\alpha/2} \sqrt {\textrm{var}\{\widehat \mu(t,x)\}}$, using  normal distributional assumption for the estimator $\widehat \mu(t,x)$, where $z^*_{\alpha/2}$ is the $100(1-\alpha/2)$ percentile of the standard normal. A robust alternative is obtained by using pointwise $100(\alpha/2)\%$ and $100(1-\alpha/2)\%$ quantiles of the bootstrap estimates $\{\widehat \mu^{(b)}(t,x): b=1,...,B \}$.

In most cases, it makes more sense to study the variability of $\widehat{\mu}(t,x)$, and draw inference about the entire true mean function $\{\mu(t,x):(t,x) \in \mathcal{D}_t \times \mathcal{D}_x\}$. Thus, we focus our study on constructing a joint (or simultaneous) confidence band for $\mu(t,x)$. Constructing simultaneous confidence bands for univariate smooths has already been discussed in the nonparametric literature. For example, \citet{degras2009simultaneous}, \citet{ma2012simultaneous}, and \citet{cao2012} proposed an asymptotically correct simultaneous confidence bands using different estimators, for independently sampled curves; %\citet{ma2012simultaneous} and \citet{cao2012} also obtain asymptotically correct pointwise confidence intervals;
\citet{CMC2011} proposed bootstrap-based joint confidence bands for univariate smooths in the case of functional data with complex error processes by using ideas of \citet{ruppert2003semiparametric}. Here, we present an extension of the approach considered by \citet{CMC2011} to bivariate smooth function.     
 
Let $\mathbf{T^*} = \{t_{g_t}: g_t=1,\ldots,G_t\}$ and  $\mathbf{X^*}=\{x_{g_x}: g_x=1,...,G_x\}$ be evaluation points that are equally spaced in the domains $\mathcal{D}_t$ and $\mathcal{D}_x$, respectively. Then, we evaluate the bootstrap estimate $\widehat{\mu}^{(b)}(t,x)$ of one bootstrap sample at all pairs  $(t, x) \in \mathbf{T^*} \times \mathbf{X^*}$, and denote by $\boldsymbol{\widehat{\mu}^{(b)}}$ the $G_tG_x$-dimensional vector with components $\widehat{\mu}^{(b)}(t,x)$. Let $\mathbf{B}$ be the $ dim(\boldsymbol{{\beta}})\times G_tG_x $-dimensional matrix obtained by column-stacking $\mathbf{B}(t_{g_t},x_{g_x})$ for all $g_t$ and $g_x$.
Let $s(t_{g_t},x_{g_x} )=\sqrt {\textrm{var}\{\widehat \mu(t_{g_t},x_{g_x})\}}$ as defined above. %and let $\boldsymbol{v}$ be the $G_tG_x $-dimensional vector of $v(t_{g_t},x_{g_x} )$ for all $g_t$ and $g_x$.
 After adjusting for the bivariate structure of the problem the main steps of the construction of the joint confidence bands for $\mu(t,x)$ follow similarly to the ones used in \citep{CMC2011} for univariate smooth parameter functions. For completeness we describe the steps below.\\

\noindent \textit{Step 1.} Generate a random variable $\mathbf{u}$ from the multivariate normal distribution with mean ${\boldsymbol{0}}_{ dim(\boldsymbol{{\beta}})}$ and variance-covariance matrix $V_{\boldsymbol{{\widehat \beta}}}$; let  $q(t_{g_t},x_{g_x})=  \mathbf{B}(t_{g_t},x_{g_x})^T \mathbf{u}$ for ${g_t}=1, \ldots, G_t$ and ${g_x}=1, \ldots, G_x$.\\

\noindent \textit{Step 2.} Calculate $q_{max}^* = \text{max}_{(t_{g_t},x_{g_x})} \text{ }\{ |q(t_{g_t},x_{g_x}) |/\sqrt{s(t_{g_t},x_{g_x})}: (t_{g_t},x_{g_x}) \in \mathbf{T^*} \times \mathbf{X^*} \}$. \\

\noindent \textit{Step 3.} Repeat \textit{Step 1.} and \textit{Step 2.} for $r = 1,\ldots ,R$, and obtain $\{q_{max,r}^*: r =  1,\ldots ,R\}$. Determine the $100(1-\alpha)\%$ empirical quantile of $\{q_{max,r}^*: r =  1,\ldots ,R\}$, say $\widehat{q}_{1-\alpha}$.\\

\noindent \textit{Step 4.} Construct the $100(1-\alpha)\% $ joint confidence band by: $\{\bar{{\mu}}(t_{g_t},x_{g_x})  \pm \widehat{q}_{1-\alpha} \ \sqrt{s(t_{g_t},x_{g_x})}:(t_{g_t},x_{g_x}) \in \mathbf{T^*} \times \mathbf{X^*} \}$. Here $\bar{{\mu}}(t,x) = B^{-1} \sum_{b=1}^B \widehat \mu^{(b)}(t,x)$ is the sample mean of the bootstrap estimates $ \widehat \mu^{(b)}(t,x)$'s.\\

\noindent The performance of the joint confidence bands is evaluated via simulation study in Section \ref{sec:simulation}. The joint confidence band provides a information about the entire true mean function. Moreover, the joint confidence band, in contrast to the pointwise confidence band, can be used as an inferential tool for formal global tests about the mean function, $\mu(t,x)$. For example, one can use the joint confidence band for testing the null hypothesis, $\text{H}_0:\mu(t,x) = 0 \text{ for all pairs } (t,x) \in \mathcal{D}_t \times \mathcal{D}_x$, by checking whether the confidence band $\{\bar{{\mu}}(t_{g_t},x_{g_x})  \pm \widehat{q}_{1-\alpha} \ \sqrt{s(t_{g_t},x_{g_x})}:(t_{g_t},x_{g_x}) \in \mathbf{T^*} \times \mathbf{X^*} \}$ contains the vector $\mathbf{0}_{G_tG_x}$. If the confidence band does not contain $\mathbf{0}_{G_tG_x}$, then we conclude that there is significant evidence that the true mean function is nonzero. Furthermore, one can use this approach to study hypothesis testing that the mean function $\mu(t,x)$ is equal to some specified bivariate smooth function, say $f_0(t,x)$, by simply investigating whether the specified function is contained in the joint confidence band.

\section{Hypothesis testing for $\mu(t,x)$}\label{sec:testStat}

Next, we focus on assessing the effect of the covariate of interest $X$ on the mean function. Consider the general case when the model is (\ref{model:initial}) and the average effect is an unspecified bivariate smooth function, $\mu(t,x)$. One of the goals is to test if the true mean function depends on $x$, that is testing the following null hypothesis: 	
	
\begin{equation} \label{eq:hypothesis}
\left.
  \begin{array}{ll}
    \text{H}_0 :& \mu(t,x) = \eta(t) \text{ for all } t,x, 
  \end{array}
  \right.
\end{equation}
for some \textit{unknown} smooth function $\eta: \mathcal{D}_t\rightarrow R$ against the alternative $\text{H}_A : \mu(t,x)$ varies over $x$ for some $t$.

To the best of our knowledge, this type of hypothesis has not been studied in functional data analysis. The problem was extensively studied in nonparametric smoothing, where the primary interest centered on significance testing of a subset of covariates in a nonparametric regression model. For example, \citet{fan1996consistent} and \citet{lavergne2000nonparametric} proposed consistent, kernel-based test statistics.  \citet{delgado2001significance} and \cite{gu2007bootstrap} also considered similar test statistics, but proposed bootstrap methods to approximate the null distribution of the test statistic. \citet{hall2007nonparametric} proposed a cross-validation based method. However, all these methods are based on the assumption that observations are independent across sampling units; in our context requiring independence of $Y_{ij}(t_{ijk})$ over $j$ and $k$ is unrealistic. Failing to account for this dependence leads to inflated type I error rates. %Nov 2014: The statement needs to be checked numerically !!

To test hypothesis (\ref{eq:hypothesis}), we propose a test statistic based on the $L^2$ distance between the mean estimators under the null and alternative hypotheses. Specifically we define the test statistic as:
\begin{equation} \label{eq:testStatistic}
T = \int_\mathcal{X} \int_\mathcal{T} \{\widehat{\mu}_A(t,x)-\widehat{\mu}_0(t) \}^2 dtdx,
\end{equation}
where $\widehat{\mu}_0(t)$ and $\widehat{\mu}_A(t,x)$ are the estimates of $\mu(t,x)$ under the null and alternative hypothesis, respectively. In particular, $\widehat{\mu}_A(t,x)$ is estimated as in Section \ref{sec:model}. The estimator $\widehat{\mu}_0(t)$ is obtained by modeling $\mu(t) = \sum_{l=1}^{d_t} B^{t}_l(t)  \beta_{l}={\mathbf{B}(t)^T} {\boldsymbol{\beta}}$ for the $d_t$-dimensional vector ${\boldsymbol{\beta}}$ and by estimating the mean parameters ${\boldsymbol{\beta}}$ based on a criterion similar to (\ref{eq:objectiveFn}). Specifically, we use the penalized criterion
%\begin{equation} \label{eq:objectiveFnUniv}
$ %\underset{\boldsymbol{\beta}, \text{ }\boldsymbol{\tau}, \text{ } \lambda_t^u}{\text{minimize}} \text{  }
 \sum_{i, j, \ell} \{Y_{ij}(t_\ell)-\mathbf{B}(t_\ell)^T\boldsymbol{\beta}-\mathbf{Z}_{ij} \boldsymbol{\tau}\}^2 + \lambda_t \boldsymbol{\beta}^T \mathbf{P_t} \boldsymbol{\beta},
$ %\end{equation}
where $\lambda_t$ is the smoothing parameter and $ \mathbf{P_t}$ is the $d_t\times d_t$ penalty matrix described in Section \ref{sec:model}. In practice, the two estimated effects $\widehat{\mu}_0(t)$ and $\widehat{\mu}_A(t,x)$ can be obtained using the \verb|gam| function in the \verb|R| \citep{R} package \verb|mgcv| \citep{wood2006}. % to implement univariate penalized spline smoothing \citep{wood2006} for estimating the mean function $\mu(t)$.  and \verb|s| 

Deriving the null distribution of the test statistic $T$ is challenging. We propose to approximate the null distribution of T using either of subjects or of subject-level residuals. Below we provide the details.

\begin{comment}
Fit the model (\ref{model:initial}) with the mean structure of interest modeled as in the alternative hypothesis $\mu(t,x)= \mathbf{B}(t,x)^T \boldsymbol{{\beta}}$; obtain the estimated mean function $\widehat \mu_A(t,x)$ by employing the estimation criterion (\ref{eq:objectiveFn}). Calculate the pseudo-residuals by $e_{ij}(t_\ell) = Y_{ij}(t_\ell) -\widehat \mu_A(t,x)-\mathbf{Z}_{ij} \widehat \tau_A 
%- \mathbf{Z}^T_{ij}\widehat{\boldsymbol{\tau}}_A
$, where $\widehat \tau_A $ is the estimate of the nuisance parameter under the alternative hypothesis. Similarly fit the null model and obtain estimates $\widehat \mu_0(t)$. Let $T_{obs}$ be the value of the test statistic using expression (\ref{eq:testStatistic}) and numerical integration. Throughout this section we suppress the dependence of the parameter estimates on smoothing parameters for simplicity of exposition.
\end{comment}

Below we provide the details of the algorithm.
\begin{algorithm}[H]
\caption{Bootstrap approximation of the null distribution of the testing procedure} 
\label{sec:bootstrapResidT0}
\begin{algorithmic}[1]
\For{$b \in \{1, \ldots, B\}$}
    \State Re-sample the subject indexes from the index set $\{1,\ldots,n\}$ with replacement. Let $I^{(b)}$ be the obtained sample of subjects. For each $i=1, \ldots, n$ denote by $m^*_i$ the number of repeated time-visits for the $i$th subject selected in $I^{(b)}$.     
    \State Define the $b$th bootstrap sample of pseudo-residuals \\
    [] $\{e^*_{ij}(t_\ell):i=1, \ldots, n, \text{ } j = 1,\ldots, m^*_{i}, \text{ and } \ell=1,\ldots,L \}$.
    For each $i=1, \ldots, n$ let $\{\mathbf{Z}^*_{ij}: j=1, \ldots, m^*_i\}$ the corresponding sample of the
    nuisance covariates for the $i$th subject selected in $I^{(b)}$. Similarly define $X^*_{ij}$.
    \State Define the $b$th bootstrap data by: \\
    [] $\text{{data}}^{(b)} =[\{Y^*_{ij}(t_\ell), X^*_{ij}, \mathbf{Z}^*_{ij}] :i=1, \ldots, n, j = 1,\ldots, m^*_{i}, \ell=1,\ldots,L \}$, where $ Y^*_{ij}(t_\ell)
    = \widehat \mu_0(t_\ell)+ \mathbf{Z}^*_{ij} \widehat \tau_A   +  e^*_{ij}(t_\ell)  $
    \State  Using $\text{{data}}^{(b)}$ fit two models. First, fit model (\ref{model:initial}) with the mean structure modeled by $\mu(t,x)= \mathbf{B}(t,x)^T \boldsymbol{{\beta}}$ and estimate $\widehat \mu^{(b)}_A(t,x)$. Second, fit model (\ref{model:initial}) with the mean structure modeled by $\mu(t)= \mathbf{B}(t)^T \boldsymbol{{\beta}}$ and estimate $\widehat \mu^{(b)}_0(t)$. Calculate the value of the test statistic $T^{(b)}$ using formula (\ref{eq:testStatistic}). 
\EndFor
\State Approximate the tail probability $P(T>T_{obs})$ by the $p\text{-value}=B^{-1}\sum_{b=1}^{B}I(T^{(b)}>T_{obs})$, where $T_{obs}$ is obtained using the original data and $I$ is the indicator function.		
\end{algorithmic}
\end{algorithm}

When the covariates $X_{ij}$ and $\mathbf{Z}_{ij}$ do not depend on visit, i.e. $X_{ij}=X_i$ and $\mathbf{Z}_{ij}=\mathbf{Z}_{i}$, the algorithm can be modified along the lines of the `bootstrap of the subject-level residuals' algorithm. %In the simulation experiment we used the above algorithm to calculate the approximated p-value. 

\section{Application to physical activity data} \label{sec:blsa}
Physical activity measured by wearable devices such as accelerometers provides new
insights into the association between activity and health outcomes \citep{Schrack:14}; The complexity of the data also poses serious challenges to current statistical analysis. For example,
accelerometers can record activity at minute level resolution
 for many days and for hundreds of individuals. Here we consider the physical activity data from the 
 Baltimore Longitudinal Study on Aging \citep{Stone:66}.  Participants in the study
wore the Actiheart portable physical activity monitor (\citealt{Brage:06}) 24 hours a day for a number of days. Activity counts
were measured in 1-min epochs and each daily activity profile has 1440 minute-by-minute measurements
of activity counts. Activity counts are proxies of activity intensity. 
Activity counts were log-transformed (more precisely, $x \rightarrow \text{log}(1+x)$) because they are highly skewed and then averaged in 5-min intervals. For simplicity, hereafter we refer to the log-transformed counts as log counts.
 For this analysis, we focus on 1779 daily activity profiles from a single visit of 
 378 female participants who have at least two days of data.
 Women in the study are aged between 31 and 93 years old. 
 Further details on the BLSA activity data can be found in \citet{Schrack:14} and \citet{xiao2014quantifying}. 

\begin{figure}[htp]
\centering
\includegraphics[height=3.25in,angle=0]{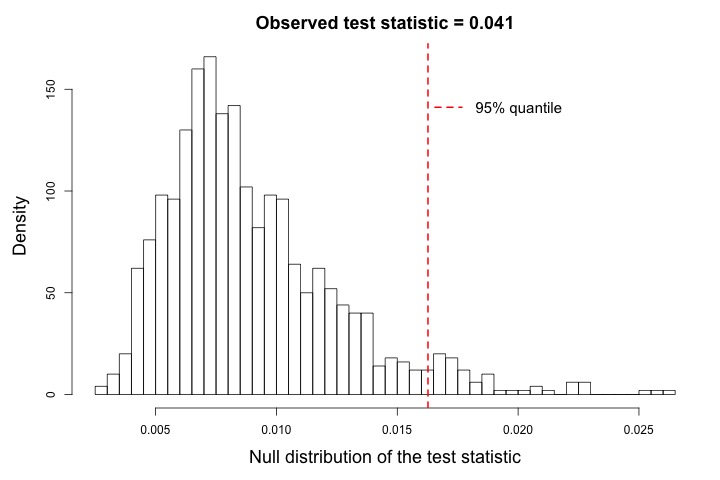}
\caption{\label{fig:hist}The null distribution of the test statistic in \eqref{eq:testStatistic} for the null hypothesis that there is no effect of age on activity. The red dashed line is the 95 percent
quantile of the null distribution of the test statistic. }
\end{figure}

Our objective is to conduct inference on the marginal effect of age on women's daily activity after adjusting for body mass index. 
We model the mean log counts as $\mu(t, X_i) + Z_i \beta(t)$, 
where $X_{i}$ and $Z_i$ are the age and body mass index of the $i$th woman during the visit, $\mu(t, x)$ is the baseline mean log counts for time $t$ within the day for a woman aged $x$ years old, and $\beta(t)$ is the association of body mass index with mean log counts for time $t$ within the day.
 We test whether $\mu(t,x)$ varies solely with $t$. We use the proposed testing statistic, $T=\int \int \{\widehat \mu_A(t,x) - \widehat \mu_0(t)\}^2 dt dx$ as detailed in Section \ref{sec:testStat}. The estimate $\widehat \mu_A(t, x)$ is based on the tensor product of $15$ cubic basis functions in $t$ and $5$ cubic basis functions in $x$ and the estimate $\widehat \mu_0(t)$ is based on $15$ cubic basis functions. Figure \ref{fig:hist} shows the null distribution. The observed test statistic is $T=0.041$ and the corresponding p-value is less than $0.001$ based on $1000$ MC samples. This indicates that there is strong evidence that daily activity profiles in women vary with age. 

\begin{figure}[htp]
\centering
\includegraphics[height=3.25in,angle=0]{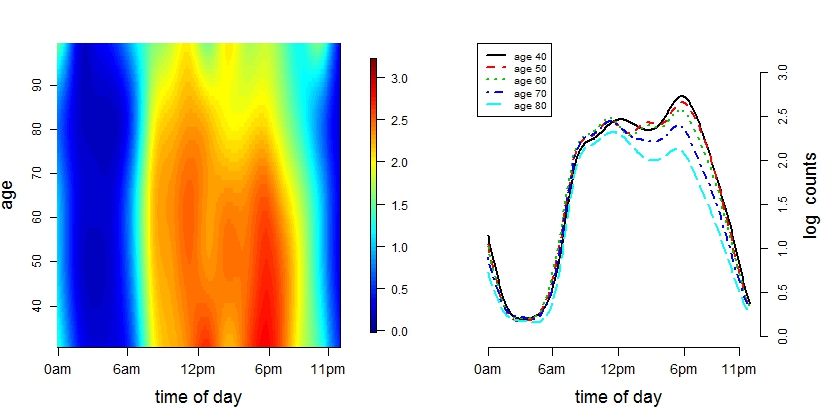}
\caption{\label{fig:mean} Heat map of average of bootstrap estimates of 
  log counts as a bivariate function of time of day and age (left panel) and
average of bootstrap estimates of log counts for five different age groups (right panel). }
\end{figure}

\begin{figure}[htp]
\centering
\includegraphics[height=3.25in,angle=0]{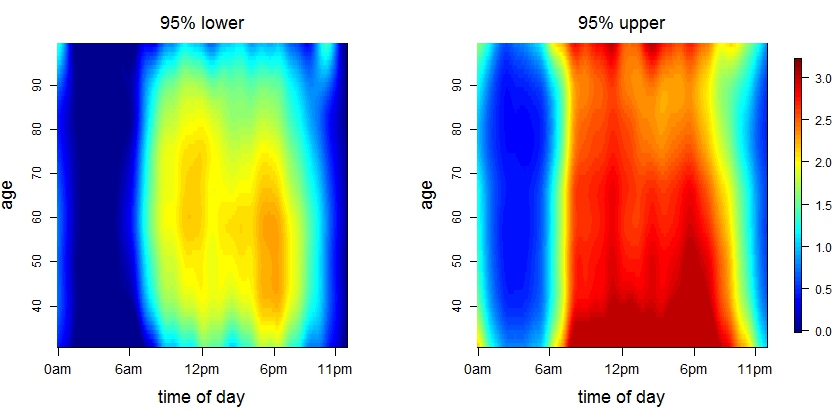}
\caption{\label{fig:confidence} Heat maps of joint confidence bands for the estimate in the left panel of Figure \ref{fig:mean}. The legend on the right applies to both plots. }
\end{figure}

\begin{figure}[htp]
\centering
\includegraphics[height=3.25in,angle=0]{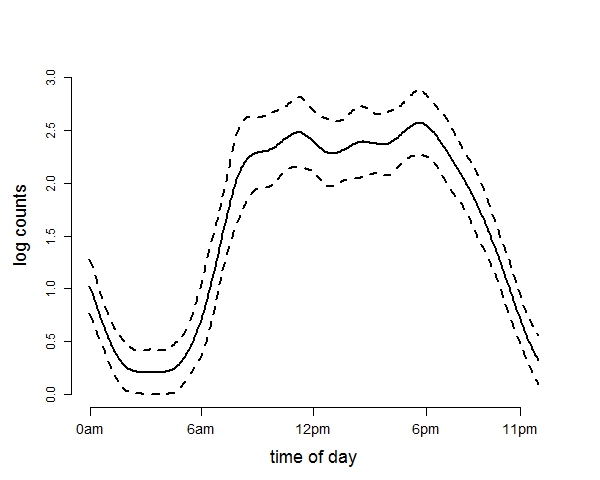}
\caption{\label{fig:age60} Average of bootstrap estimates of log counts as a function of time of day at age 60 and the associated joint confidence bands. }
\end{figure}

\begin{figure}[htp]
\centering
\includegraphics[height=3.25in,angle=0]{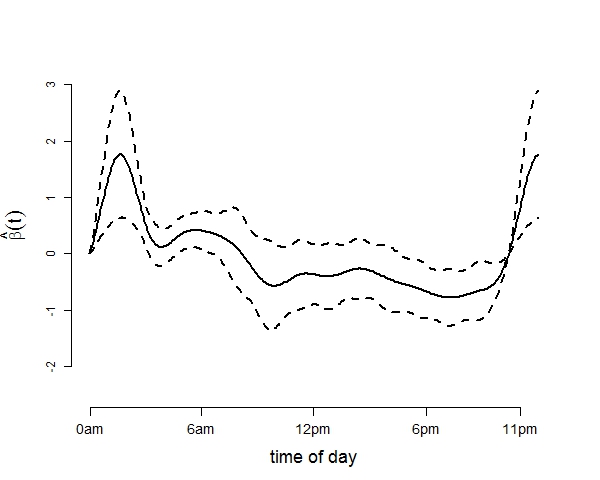}
\caption{\label{fig:bmi} Association of body mass index with mean log counts as  a function of time of day and the associated joint confidence bands. }
\end{figure}

Figure \ref{fig:mean} shows the  estimated baseline activity profile as a function of age, $\widehat \mu(t,x)$, average of all bootstrap estimates.  The plot indicates that the average log counts is a decreasing function of age for most time during the day. Furthermore, it depicts two activity peaks, one around 12pm and the other around 6pm. In particular, the peak in the evening seems to decrease faster with age, indicating that afternoon activity is more affected by age. The joint lower and upper $95\%$ confidence limits based on methods described in Section \ref{sec:variability} are displayed in Figure \ref{fig:confidence}. Figure \ref{fig:age60} displays the estimated mean activity profile for 60 years old women along with the corresponding joint $95\%$ confidence band. Figure \ref{fig:bmi} displays the estimated association of body mass index with mean log counts as a function of time of day; it suggests that women with higher body mass index have less activity during the day and evening, albeit more activity at late night and early morning.
\\
\\
%\noindent \textcolor{blue}{\textbf{Questions for LUO:} \\
%(0) Fig 2 right hand plot looks awfully familiar to the one presented in your other paper. If so, let's move it to the supplementary material. Could you please look into this ? We should only include pictures that have not yet included in previously appeared papers in the main ms. We can include them for completeness in the Appendix. Please make the necessary changes and update this section. Thanks. \\
%(1) NEED information about tuning parameter, which basis functions are used, the basis dimensions, knots sequence selection\\
%(2) SHALL we include results for joint confidence band ?\\
%%(3) descriptive plot of the activity profiles for female participants only and/or plot of the estimated bivariate mean function from \citet{xiao2014quantifying} for the Supplementary Material}\\
%\\
%\textcolor{red}{Please clarify Figure 2  - my understanding and updated description seems to be different from the label of this plot. Could you please clarify, and make the appropriate changes ?
%%in the main manuscript for now, as I realized that it is average of bootstrap-estimates of a bivariate mean function - from my understanding, this is a different plot from the previous paper.
%}}
%\clearpage

\FloatBarrier
\subsection{Validating the testing results via simulation study}

We conducted a simulation study designed to closely mimic the BLSA data structure. Specifically we generate data from model (\ref{model:initial}) with $\mu(t,x) = \text{cos}(2\pi t) + \delta(\widehat{\mu}(t,x) - \text{cos}(2\pi t))$, where $ \widehat{\mu}(t,x) $ is the estimated mean log counts, $\delta$ is some parameter quantifying the distance from the null and alternative hypotheses, $\boldsymbol{\tau} = \textbf{0}$ (i.e. there is no additional covariate vector), and the errors $\epsilon_{ij}(t)$ are generated to have a covariance structure that mimics that of residuals from the BLSA data \citep{xiao2014quantifying}. Notice that when $\delta=0$ the true mean profile $\mu(t,x) =\text{cos}(2\pi t) $, whereas when $\delta =1$ then $\mu(t,x)=\widehat{\mu}(t,x) $. The covariate $X_{i}$ and the number of visits per subject, $m_{i}$, are generated uniformly from $\{30,\ldots,90\}$ and $\{5,\ldots,9\}$ respectively. We use $n=378$, the number of female participants in the BLSA. 

\begin{figure}[ht!] %\label{figSup: power}
   \centering
 \includegraphics[scale=0.45]{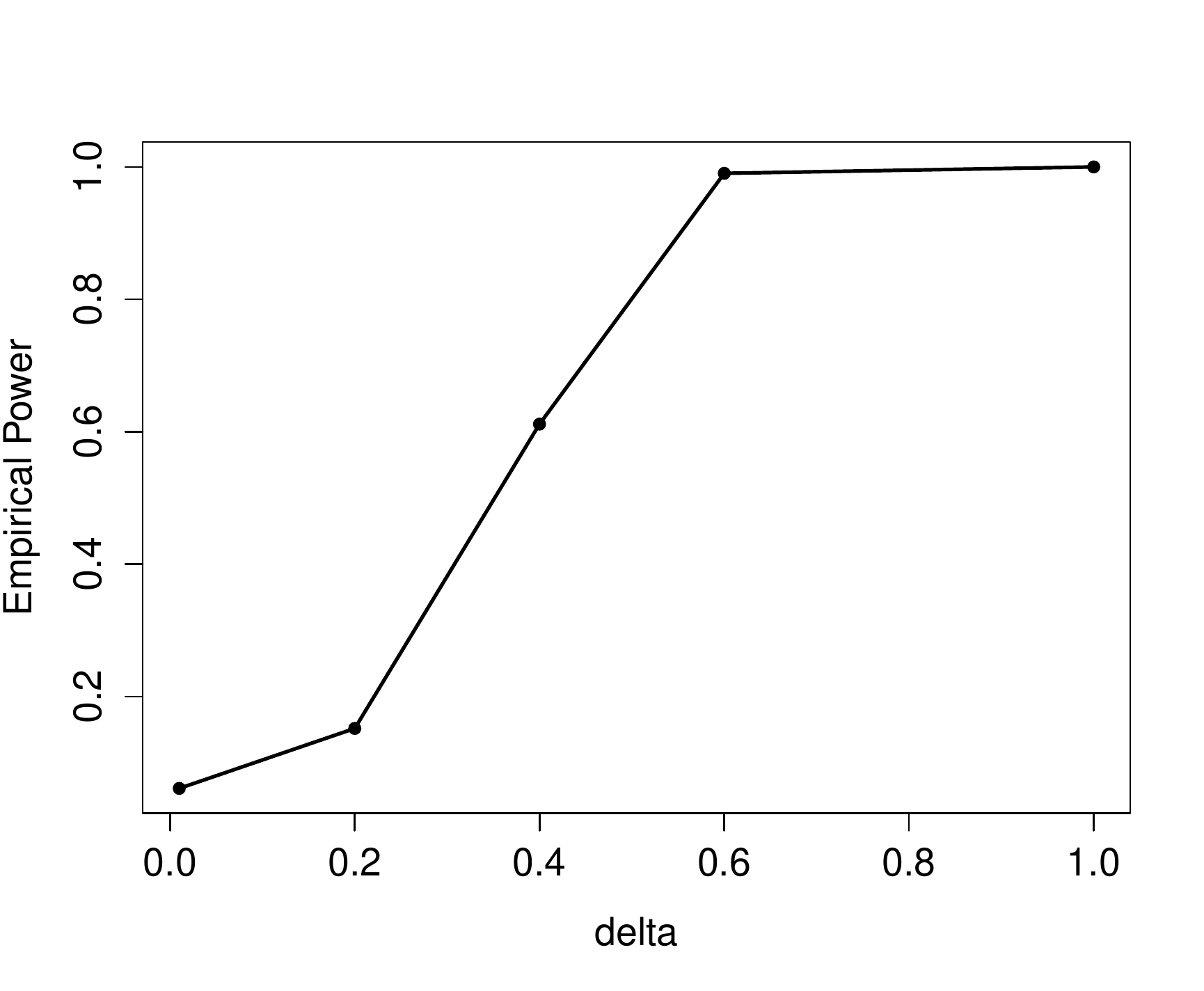} %{powerBLSAsim.png}
    \caption{Estimated power curves for testing $H_0: \mu(t,x)=\eta(t)$ using $\alpha = 0.05$, when the true mean function $\mu(t,x)=\text{cos}(2\pi t) + \delta(\widehat{\mu}(t,x) - \text{cos}(2\pi t))$ for $\delta=0.01,2,4,6, 8$. Results are based on $N_{sim}=500$ MC samples.}
  \label{figSup:PowerBLSAsim}
   \end{figure}

Table \ref{tabSup: size} shows the rejection probabilities in 1000 simulations, when $\delta=0$ and indicates that the empirical Type I error of the proposed testing procedure is close to the nominal level. 
Figure \ref{figSup:PowerBLSAsim} displays the rejection probabilities in 500 simulations, when $\delta>0$. For all cases, we use $B=300$ bootstrap samples to approximate the null distribution of the test statistic $T$.  
%\clearpage
\begin{table}[!ht]
\caption{Empirical type I error of the test statistic $T$ based on the $N_{sim}=1000$ MC samples; Mean function is $\mu(t,x)=\text{cos}(2\pi t)$, $\tau=0$}
\centering
\scalebox{0.85}{
\begin{tabular}{ccc}\hline   \label{tabSup: size}
% \multicolumn{3}{c}{$\mu(t,x)=\text{cos}(2\pi t)$, $\tau=0$}\\ \hline
  $\alpha=0.05$ & $\alpha=0.10$ & $\alpha=0.15$\\ \hline
 0.06 & 0.11 & 0.16 \\ 
(0.01) & (0.01) & (0.01) \\ \hline
\multicolumn{3}{l}{\scriptsize Standard errors are presented in parentheses.}
 \end{tabular}}
\end{table}

\vspace{-1em}

\FloatBarrier

\section{Simulation Study} \label{sec:simulation}

In this section we evaluate the performance of the inferential methods introduced in this paper. First, we evaluate the accuracy of the pointwise and joint confidence bands in terms of average coverage probability and average confidence interval length. Second, we evaluate the testing procedure with respect to Type I error and power. 

Data are simulated using the model (\ref{model:initial}) where $X_{ij} = X_{i}$, $Z_{ij}=Z_i$. Errors $\epsilon_{ij}(t)$ are generated from $\epsilon_{ij}(t) = \sum_{l=1}^{3} \xi_{ijl}\phi_l(t) + w_{ij}(t)$,
%\begin{equation} \label{eq:simulationModel}
%\epsilon_{ij}(t) = \sum_{l=1}^{3} \xi_{il}(T_{ij})\phi_l(t) + w_{ij}(t)
%\end{equation}
%\noindent for $i=1,\ldots,n$, $j=1,\ldots,m$, and $t \in \mathcal{D}_t = [0,1]$, 
where $\xi_{ijl}$ are random variables with zero mean, variance $\lambda_l$ that are independent over $i$ and $l$, and exponential autocorrelation with a correlation parameter $\rho$. The residuals $w_{ij}(\cdot)$ are mutually independent with zero mean and variance $\sigma^2$. The number of repeated measures is fixed at $m_i=5$, $(\lambda_1, \lambda_2, \lambda_3)=(3,2,1/3)$, and the functions $[\phi_1(t),\phi_2(t),\phi_3(t)]=[\sqrt{2}\text{cos}(2\pi t),\sqrt{2}\text{sin}(2\pi t), \sqrt{2}\text{cos}(4\pi t)]$. The subject-specific covariates $X_i$ and $Z_i$ are generated from a Uniform$[0,1]$. The grid of points $\{t_\ell: \ell=1, \ldots, L\}$ is set as $101$ equally spaced discrete points in $[0,1]$. The variance of the white noise process $\sigma^2$ is set to $5.33$, which is equivalent to ensuring a signal to noise ratio equal to $1$. Here the signal to noise ratio is calculated as SNR$=\int\text{var}[Y_{ij}(t)]dt/\sigma^2 - 1 = \sum_{l=1}^{3}\lambda_l/\sigma^2$. \\
We consider different combinations of the following factors: 
\begin{enumerate} [label=F\arabic*]
\item number of subjects: (a) $n=100$, (b) $n=200$, and (c) $n=300$;
\item mean function: \label{simFactor:MeanFn}

\begin{tabular}{rl} 
 \ref{simFactor:MeanFn} i. & bivariate function $\mu(t,x)$ \\
(a) & $\mu_1(t,x) = \beta_0 + \beta_tt + \beta_xx$ for $(\beta_0, \beta_t, \beta_x) = (5,2,3)$,(Ex \ref{ex:linear}) \\
(b) &  $\mu_2(t,x) = \beta_0 + \beta_tt + \beta_xx+\beta_{tx}tx$ for $(\beta_0, \beta_t, \beta_x, \beta_{tx}) = (5,2,3,7)$, (Ex \ref{ex:interaction})\\
(c) & $\mu_3(t,x)= \text{cos}(2\pi t) + \beta_xx \text{ for } \beta_x = 3$, (Ex \ref{ex:univariate})\\ 
(d) & $\mu_4(t,x)= \text{cos}(2\pi t) + \delta((x/4)-t)^3 \text{ for }\delta=0,2,4$, and $6$, (Ex \ref{ex:nonparametric}) \\ 
 \ref{simFactor:MeanFn} ii. & linear effect of covariate $Z_i$\\
(a) & $\tau=0$ (no effect), (b) $\tau=8$;
\end{tabular}

\item correlation:
(a) $\rho = 0.2$ (weak correlation) and 
(b) $\rho=0.9$ (strong correlation).
%\item signal-to-noise ratio (SNR), where SNR$=\int\text{var}[Y_{ij}(t)]dt/\sigma^2 - 1 = \sum_{l=1}^{3}\lambda_l/\sigma^2$: \\
%(a) relatively small noise; SNR $=1$,\\ 
%(b) relatively large noise; SNR $=0.25$. 
\end{enumerate}

%  average coverage probability as well as average length
Our methodology is evaluated on these models in two ways. First, we model the data by assuming the correct model and by evaluating the accuracy of the inferential procedures; the results are detailed next. Second, we model the data using a bivariate mean, $\mu(t,x)$, and evaluate the performance of the confidence bands of $\mu(t,x)$ for covering the true mean even when the true mean has a simpler structure; results are described in the Supplementary Material section S3.

We show now results for fitting the correct model; estimation is done as detailed in Section \ref{sec:model}. When the assumed model for the mean structure of interest involves univariate or bivariate smooths, we use $d_t=7$ and/or $d_x=7$ cubic B-spline basis functions, and select the smoothing parameter/s via GCV; specifically for the bivariate smooth, $d_td_x = 49$ basis functions are used. Compared to the data analysis, we use a relatively small number of basis functions because there are only $101$ grid points along $t$. Estimation accuracy is measured using the bias and variance of the estimators; for univariate and bivariate smooths, single number summaries of these measures are used. Specifically, when the mean of interest is $\mu(t)=\cos(2\pi t)$, as in scenario F2 i.(c), the integrated bias defined by $\int_0^1 \{  \bar \mu(t)  - \mu(t)\} dt $ is used as a summary measure of bias, and the integrated variance, defined by $\int_0^1 \{\sum_{i_{sim}=1}^{ N_{sim}}  \{  \widehat \mu_{i_{sim}}(t) - \bar \mu(t) \}^2 / (N_{sim}-1) \} dt $ is used as a summary measure of variance. Here $\widehat \mu_{i_{sim}}(t)$ denotes the mean estimator from one simulation,  $\bar \mu(t) =  \sum_{i_{sim}=1}^{ N_{sim}}\widehat \mu_{i_{sim}}(t)/ N_{sim}$ is the sample mean of the estimator $\widehat \mu(t)$. 
Inference for the parameter/s of interest is done using methods described in Sections \ref{sec:variability} and \ref{sec:testStat}. The performance of the pointwise and joint confidence bands for both univariate, and bivariate cases is evaluated in terms of average coverage probability (ACP), and average length (AL). Specifically, let $(\widehat \mu^{i_{sim},\ l}(t,x), \ \widehat \mu^{i_{sim},\ u}(t,x))$ be the $100(1-\alpha)\%$ pointwise confidence interval of $\mu(t,x)$ obtained at the $i_{sim}$ Monte Carlo generation of the data, then   

$$
\text{ACP}^{\text{point}} = \frac{1}{N_{sim}G_tG_x} \sum_{i_{sim}=1}^{N_{sim}} \sum_{g_t=1}^{G_t} \sum_{g_x=1}^{G_x} 1\left \{ \mu(t_{g_t},x_{g_x}) \in (\widehat \mu^{i_{sim},\ l}(t_{g_t},x_{g_x}), \ \widehat \mu^{i_{sim},\ u}(t_{g_t},x_{g_x})) \right \} 
$$
$$
\text{AL}^{\text{point}} = \frac{1}{N_{sim}G_tG_x} \sum_{i_{sim}=1}^{N_{sim}} \sum_{g_t=1}^{G_t} \sum_{g_x=1}^{G_x} |\widehat \mu^{i_{sim},\ l}(t_{g_t},x_{g_x})- \ \widehat \mu^{i_{sim},\ u}(t_{g_t},x_{g_x}))|,
$$
where $ \{t_{g_t}: g_t=1,\ldots,G_t\}$ and  $\{x_{g_x}: g_x=1,...,G_x\}$ are equi-distanced grid points in the domains $\mathcal{D}_t$, and $\mathcal{D}_x$, respectively. Next, let $(\widehat \mu^{i_{sim},\ l}(t,x), \ \widehat \mu^{i_{sim},\ u}(t,x))$ be $100(1-\alpha)\%$ joint confidence interval. The average length is calculated as above, while the average coverage probability is calculated as:
$$
\text{ACP}^{\text{joint}} = \frac{1}{N_{sim}} \sum_{i_{sim}=1}^{N_{sim}} %\sum_{g_t=1}^{G_t} \sum_{g_x=1}^{G_x} 
1\left \{ \mu(t_{g_t},x_{g_x}) \in (\widehat \mu^{i_{sim},\ l}(t_{g_t},x_{g_x}), \ \widehat \mu^{i_{sim},\ u}(t_{g_t},x_{g_x})) : \text{ for all } g_t, g_x \right \} 
%: g_t=1,\ldots,G_t, g_x=1,\ldots,G_x \right \}. 
$$

The performance of the test statistic $T$ is evaluated in terms of its empirical type I error (size) for the nominal levels of $0.05$, $0.10$, and $0.15$, and power for the nominal level of $0.05$. The results for the nominal coverage of $95\%$ are presented in Table \ref{tab:JCB}; the results for other nominal coverages ($85\%$ and $90\%$) are included in the Supplementary Material. 

The results for the empirical size of the testing procedure are based on $N_{sim}=1000$ MC samples, while the results for the coverage probability, expected length, and power of the test are based on $N_{sim}=500$ MC samples. For each MC simulation we use $B=300$ bootstrap samples; they are obtained by bootstrapping residuals by subject. %The relevant results based on bootstrapping subject-level observations are included in the Supplementary Material section \ref{SUPsection: bootstrap subject-level observations}.  

Table \ref{tab:JCB} shows the bias, variance, ACP and AL for the mean structure of interest and using the nominal level $95\%$ when the sample size is $n=100$. When the mean structure includes smooth terms, both pointwise and joint confidence intervals/bands are provided. Overall, both pointwise or/and joint confidence intervals/bands perform well. The confidence interval/bands tend to be wider when the correlation within the repeated observations is strong ($\rho=0.9$) than when is weaker ($\rho=0.2$). %We also consider small SNR ($\text{SNR}=0.25$) for different true mean functions, and the results (not presented here) show that signal-to-noise ratio (SNR) does not affect empirical coverage, and average integrated length of a confidence band. 

The joint confidence bands based on bootstrap of subjects perform equally well when the effect of the covariate $X$ is linear (cases F2 i.(a)-(c)). For the case of the nonlinear effect of $X$ on the outcome (case F2 i.(d)), we consider both a covariate that is constant over visit (i.e. $X_i$) and a covariate that varies with visit (i.e. $X_{ij}$). The results show the good coverage of the joint confidence bands with the visit-varying covariate $X_{ij}$. Additional results based on bootstrapping subject-level observations are included in the Supplementary Material section S2. \modif{The results suggest that for a time-invariant covariate (i.e. $X_i$) the bootstrap of subject-level residuals gives a narrower joint confidence band with better coverage than the bootstrap of subject-level observations.}     
%Results based on bootstrap of subjects perform equally well when the effect of the covariate $X$ is linear (cases F2 i.(a)-(c)), but shows some undercoverage when the effect of $X$ on the outcome is nonlinear (case F2 i.(d)). The undercoverage is expected as a covariate we consider here is time-invariant. A time-varying covariate (i.e. $X_{ij}$ instead of $X_i$) is considered for the case of F2 i.(d), and the results for the two types of bootstrap are equivalent in this case. Additional results based on bootstrapping subject-level observations are included in the Supplementary Material section \ref{SUPsection: bootstrap subject-level observations}.      

% % % % % % % % % % % % % % % % % % % % % % % % % % % % % % % % % % % % % %
%
% % % % % % % % % % % % % % % % % % % % % % % % % % % % % % % % % % % % % %

\begin{figure}[ht!]
   \centering
   \includegraphics[scale=0.45]{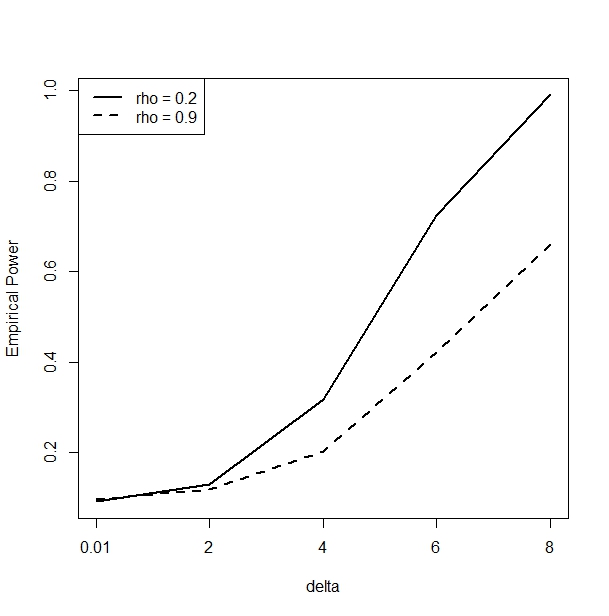}
   \caption{Estimated power curves for testing $H_0: \mu(t,x)=\eta(t)$ using level of significance $\alpha = 0.05$, when the true mean function $\mu(t,x)=2\cos (2\pi t)+ \delta (x/4-t)^3$ for $\delta=0.01,2,4,6$. The results are based on $N_{sim}=500$ MC samples.}
  \label{figure:Power}
   \end{figure}

Table \ref{tab:size} shows the empirical type I error of the proposed testing procedure for testing $H_0: \mu(t,x)=\eta(t)$, where $\eta(\cdot)$ is a smooth effect depending on $t$ only. Rejection probabilities are given for various nominal levels, different correlation strength, and increasing sample sizes. Results indicate that, as sample size increases, the size of the test gets closer to the corresponding nominal levels. Including an additional covariate in the model seems to have no effect on the performance of the testing procedure. Figure \ref{figure:Power} illustrates the power curves, when the true mean structure deviates from the null hypothesis. It presents the power as a function of the deviation from the null that involves both $t$ and $x$, $\mu(t,x) = 2\cos (2\pi t)+ \delta (x/4-t)^3$. Here $\delta$ quantifies the departure from the null hypothesis. As expected, rejection probabilities increase as the departure from the null hypothesis increases, irrespective of the direction in which it deviates. As expected, rejection probabilities increase with the sample size. Our investigation indicates that the strength of the correlation between the functional observations corresponding to the same subject affect the rejection probability: the weaker the correlation, the larger the power. \modif{There is no competitive testing method available for this null hypothesis.}

The above discussion focuses on the performance of confidence intervals/bands when the correct mean structure is assumed in the estimation procedure. In the Supplementary Material section S3 we present the corresponding results when the fitted model is completely nonparametric; of course this choice is more computationally intensive. \modif{Lastly we conducted an additional simulation study to evaluate robustness of the proposed methods to the non-Gaussian error distributions and obtained the similar results as the Gaussian case; the results are presented in the Supplementary Material section S4.}

\begin{table}[ht]
\caption{Simulation results using bootstrap of subject level residuals and $95\%$ nominal level; results are based on $500$ MC samples.} 
\centering
\scalebox{0.65}{
\begin{tabular}{cllc|ccclclclcl} \hline  \label{tab:JCB}
Case& True Mean Function & Parameter & $\rho$ & Bias &
$\sqrt{\text{\textrm{var}}}$ & \multicolumn{2}{c}{$\text{ACP}^{\text{point}}$} & \multicolumn{2}{c}{$\text{AL}^\text{point}$} & \multicolumn{2}{c}{$\text{ACP}^{\text{joint}}$} & \multicolumn{2}{c}{$\text{AL}^\text{joint}$} \\ \hline 

(a) & $ \beta_0 + \beta_tt + \beta_xX +\tau Z$ & $\beta_0=5$    & 0.20 & 0.00 & 0.20 & 0.94 & (0.01) & 0.54 & (< 0.01) &  &  &  &  \\ 
&  & &      0.90 & -0.01 & 0.27 & 0.94 & (0.01) & 0.70 & (< 0.01) &  &  &  &  \\ 
&&  $\beta_t=2$ &   0.20 & 0.01 & 0.40 & 0.94 & (0.01) & 1.06 & (< 0.01) &  &  &  &  \\ 
&  &  &   0.90  & 0.01 & 0.52 & 0.94 & (0.01) & 1.39 & (0.01) &  &  &  &  \\ 
&&  $\beta_x=3$ &   0.20 & 0.00 & 0.05 & 0.95 & (0.01) & 0.14 & (< 0.01) &  &  &  &  \\ 
&  &  &   0.90  & 0.00 & 0.05 & 0.95 & (0.01) & 0.14 & (< 0.01) &  &  &  &  \\ 
&&  $\tau=8$   & 0.20 & 0.00 & 0.05 & 0.93 & (0.01) & 0.14 & (< 0.01) &  &  &  &  \\ 
&  &  &   0.90  & 0.00 & 0.05 & 0.93 & (0.01) & 0.14 & (< 0.01) &  &  &  &  \\ \hline
(b)& $ \beta_0 + \beta_tt + \beta_xX+\beta_{tx}tX  + \tau Z$ & $\beta_0=5$   & 0.20 & -0.01 & 0.39 & 0.93 & (0.01) & 1.04 & (< 0.01) &  &  &  &  \\ 
&  &  &   0.90  &-0.02 & 0.51 & 0.94 & (0.01) & 1.36 & (0.01) &  &  &  &  \\ 
&  &$\beta_t=2$   & 0.20 & 0.03 & 0.78 & 0.93 & (0.01) & 2.07 & (0.01) &  &  &  &  \\ 
&  &  &   0.90  & 0.04 & 1.02 & 0.94 & (0.01) & 2.72 & (0.01) &  &  &  &  \\ 
&  &$\beta_x=3$  & 0.20 & 0.02 & 0.67 & 0.93 & (0.01) & 1.08 & (0.01) &  &  &  &  \\ 
&  &  &   0.90  & 0.02 & 0.88 & 0.93 & (0.01) & 2.36 & (0.01) &  &  &  &  \\ 
&  &$\beta_{tx}=7$   & 0.20 & -0.04 & 1.33 & 0.92 & (0.01) & 3.60 & (0.02) &  &  &  &  \\ 
&  &  &   0.90  &-0.06 & 1.75 & 0.93 & (0.01) & 4.71 & (0.02) &  &  &  &  \\ 
&  &$\tau=8$   & 0.20 & 0.00 & 0.05 & 0.93 & (0.01) & 0.14 & (< 0.01) &  &  &  &  \\ 
&  &  &   0.90  & 0.00 & 0.05 & 0.93 & (0.01) & 0.14 & (< 0.01) &  &  &  &  \\ \hline
(c)&$\text{cos}(2\pi t)+\beta_xX +\tau Z$ & $f(t)=\text{cos}(2\pi t)$ & 0.20 & 0.00 & 0.25 & 0.93 & (0.01) & 0.67 & (< 0.01) & 0.92 & (0.01) & 0.95 & (< 0.01) \\ 
&  &  &   0.90  & 0.00 & 0.32 & 0.93 & (0.01) & 0.87 & (< 0.01) & 0.93 & (0.01) & 1.23 & (< 0.01)\\ 
&    & $\beta_x=3$ &   0.20 & 0.00 & 0.05 & 0.95 & (0.01) & 0.14 & (< 0.01) &  &   &  &   \\ 
&  &  &   0.90  & 0.00 & 0.05 & 0.95 & (0.01) & 0.14 & (< 0.01) &  &   &  &   \\ 
&    & $\tau=8$&    0.20 & 0.00 & 0.05 & 0.93 & (0.01) & 0.14 & (< 0.01) &  &   &  &  \\ 
&  &  &   0.90  & 0.00 & 0.05 & 0.93 & (0.01) & 0.14 & (< 0.01) &  &  &  &   \\ \hline
(d)&$\text{cos}(2\pi t) + 4((X/4)-t)^3+\tau Z$ & $\mu_4(t,X)$ & 0.20 & 0.00 & 0.61 & 0.94 & (< 0.01) & 1.65 & (0.01) & 0.93 & (0.01) & 3.23 & (0.01) \\ 
&  &  &   0.90  & 0.00 & 0.81 & 0.94 & (< 0.01) & 2.18 & (0.02) & 0.93 & (0.01) & 4.26 & (0.01) \\ 
&&  $\tau=8$ &   0.20 & 0.00 & 0.05 & 0.93 & (0.01) & 0.14 & (< 0.01) &  &  &  &  \\ 
&  &  &   0.90  & 0.00 & 0.05 & 0.94 & (0.01) & 0.14 & (< 0.01) &  &  &  &  \\ \hline
%(e)&$\text{cos}(2\pi t) + 4((X/4)-t)^3$ & $\mu_4(t,X)$ & 0.20  & 0.00 & 0.62 & 0.93 & (< 0.01) & 1.66 & (0.01) & 0.92 & (0.01) & 3.24 & (0.01) \\ 
%&  &  &   0.90  & 0.00 & 0.82 & 0.93 & (< 0.01) & 2.18 & (0.02) & 0.91 & (0.01) & 4.27 & (0.01) \\ \hline
\multicolumn{12}{l}{Standard errors are presented in parentheses.}
 \end{tabular}
}
\end{table}	
\FloatBarrier

\begin{table}[!ht]
\caption{Empirical Type I error of the test statistic $T$ based on the $N_{sim}=1000$ MC samples. }
\centering
\scalebox{0.85}{
\begin{tabular}{cc|cc|cc|cc}\hline  \label{tab:size}
&& \multicolumn{6}{c}{$\mu(t,x)=\text{cos}(2\pi t)$, $\tau=0$}\\ \hline
  &&\multicolumn{2}{c|}{$\alpha=0.05$}&\multicolumn{2}{c|}{$\alpha=0.10$}&\multicolumn{2}{c}{$\alpha=0.15$}\\ \hline

 $n=100$ & $\rho=0.2$ &  0.08 & (0.01) & 0.14 & (0.01) & 0.21 & (0.01) \\ 
 & $\rho=0.9$ &  0.09 & (0.01) & 0.14 & (0.01) & 0.20 & (0.01) \\ 
 $n=200$ & $\rho=0.2$ &  0.07 & (0.01) & 0.13 & (0.01) & 0.17 & (0.01) \\ 
 & $\rho=0.9$ &  0.08 & (0.01) & 0.12 & (0.01) & 0.18 & (0.01) \\ 
 $n=300$ & $\rho=0.2$ &  0.06 & (0.01) & 0.11 & (0.01) & 0.16 & (0.01) \\ 
 & $\rho=0.9$ &  0.06 & (0.01) & 0.12 & (0.01) & 0.16 & (0.01) \\ \hline
  
   && \multicolumn{6}{c}{$\mu(t,x)=\text{cos}(2\pi t)$, $\tau=8$}\\ \hline
    &&\multicolumn{2}{c|}{$\alpha=0.05$}&\multicolumn{2}{c|}{$\alpha=0.10$}&\multicolumn{2}{c}{$\alpha=0.15$}\\ \hline
  
 $n=100$ & $\rho=0.2$ &  0.07 & (0.01) & 0.15 & (0.01) & 0.20 & (0.01) \\ 
 & $\rho=0.9$ &  0.08 & (0.01) & 0.15 & (0.01) & 0.21 & (0.01) \\ 
 $n=200$ & $\rho=0.2$ &  0.07 & (0.01) & 0.13 & (0.01) & 0.17 & (0.01) \\ 
 & $\rho=0.9$ &  0.08 & (0.01) & 0.12 & (0.01) & 0.18 & (0.01) \\ 
 $n=300$ & $\rho=0.2$ &  0.06 & (0.01) & 0.11 & (0.01) & 0.16 & (0.01) \\ 
 & $\rho=0.9$ &  0.06 & (0.01) & 0.12 & (0.01) & 0.16 & (0.01) \\ \hline

\multicolumn{8}{l}{Standard errors are presented in parentheses.}
    \end{tabular}}
\end{table}

\FloatBarrier
%\clearpage

\section{Discussion} \label{sec:discussion}

In this paper we introduced statistical inference for population level effects for complex correlated functional data. We considered model fitting using conventional modeling approaches that are publicly available and computationally feasible, in particular the \verb|gam| function in \verb|R| \citep{R} package \verb|mgcv| \citep{wood2006}. Other choices may be more appropriate for model fitting in some cases: for example the sandwich estimation approach of \cite{xiao2013fast} is a much faster method to fit bi-variate smooths when the time $t$ and the covariate of interest $X_i$ are observed on a regular grid. The selection of the smoothing parameter/s using leave one-subject out cross-validation may further improve the performance of the proposed methods.

Although the fitting procedure is based on the working independence assumption, the construction of the confidence intervals as well as the testing procedure rely on the bootstrap of subjects that accounts for the complex dependence. Most importantly, the procedure we proposed is easy to implement and explain.

\section*{Acknowledgement} \label{sec:Ack}
Staicu's research was supported by NSF grant numbers
DMS 1007466 and DMS 0454942% USE DMS 1454942
and NIH grants R01 NS085211 and R01 MH086633. Crainiceanu and Xiao's research was supported by NIH grants R01 NS085211, R01 NS060910,  R01 HL123407 as well as NIA contracts HHSN27121400603P and HHSN27120400775P. Data for these analyses were obtained from the Baltimore Longitudinal Study of Aging, a study performed by the National Institute on Aging.
% % % TO ADD 

\section*{Supplementary Material} \label{sec:SM}

Additional numerical investigations are included in a supplementary material that is available online.

% % % % TABLES/FIGURES

\clearpage
\bibliography{reference}
\bibliographystyle{apalike}
%\bibliographystyle{imsart-nameyear}

%\textcolor{blue}{\textbf{note:}\\
%- The bootstrap-based testing procedure is not based on distributional assumption of the complex dependence. This has the potential extension to test for linear effect (discussion in the context of our application)
%\\
%- One-subject-out cross validation may improve the smoothing parameter selection.}

\end{document}